\documentclass[ams, prb, twocolumn, amssymb,floatfix]{revtex4-2}
\usepackage{amsmath}
\usepackage{tabularx}
\usepackage{multirow,bigstrut}
\usepackage{bm}
\usepackage{slashed}
\usepackage{euscript}
\usepackage{graphicx}
\usepackage{color}
\usepackage{amsfonts}
\usepackage{exscale}
\usepackage{amsbsy}
\usepackage{subfigure}
\usepackage{textcomp}
\usepackage{comment}
\usepackage{mathrsfs}
\usepackage{bbold}
\usepackage{blkarray}
\usepackage{ulem}

\usepackage{hyperref}

\newcommand{\beq}{\begin{equation}}
\newcommand{\eeq}{\end{equation}}
\newcommand{\be}{\begin{eqnarray}}
\newcommand{\ee}{\end{eqnarray}}

\newcommand\footnoteref[1]{\protected@xdef\@thefnmark{\ref{#1}}\@footnotemark}

\begin{document}
	\title{Numerical Study of a Dual Representation of the Integer Quantum Hall Transition}
	\author{Kevin S. Huang$^1$, S. Raghu$^{1,2}$, Prashant Kumar$^{1,3}$}
	\affiliation{$^{1}$Stanford Institute for Theoretical Physics, Stanford University, Stanford, California 94305, USA\\
	$^{2}$Stanford Institute for Materials and Energy Sciences,\\
	SLAC National Accelerator Laboratory, Menlo Park, CA 94025, USA\\
	$^{3}$Department of Physics, Princeton University, Princeton, New Jersey 08544, USA}
	\date{\today}
	
	\begin{abstract}
		We study the critical properties of the non-interacting integer quantum Hall to insulator transition (IQHIT) in a ``dual'' composite-fermion (CF) representation.  A key advantage of the CF representation over electron coordinates is that at criticality, {\it CF states are delocalized at  all} energies.  The CF approach thus enables  us to study the transition from a new vantage point. Using a lattice representation of CF mean-field theory, we compute the critical and multifractal exponents of the IQHIT. We obtain $\nu = 2.56 \pm 0.02$ and $\eta = 0.51\pm 0.01$, both of which are consistent with the predictions of the Chalker-Coddington network model formulated in the electron representation. 
	\end{abstract}

	\maketitle
	
	\noindent \textit{ Introduction - }  The integer quantum Hall (QH) to insulator transition (IQHIT) is one of the most studied topological phase transitions in condensed matter physics\cite{PhysRevLett.61.1294,Engel1993, Shahar1995,Shahar1996,Yang2000,SondhiGirvinCariniShahar}.  Without interactions, the existence of a QH plateau requires quenched disorder, and a magnetic field tunes the system  from a QH state  to an Anderson insulator.  A beautiful representation of the IQHIT known as the Chalker-Coddington model (CCM) involves  percolation of droplets of QH and insulating phases\cite{chalkercoddington}.  The CCM has been amenable to large scale numerical studies of critical exponents of the non-interacting IQHIT\cite{Huckestein1999}.

	Nevertheless, all electron representations of the IQHIT suffer with a drawback: it is difficult to include electron-electron interaction effects, which are necessary to account for very basic aspects of the IQHIT.  Interactions are necessary to ensure a non-zero finite temperature electrical resistivity\cite{Wang2000}.  Moreover, interactions determine dynamical scaling laws and superuniversality (the issue of whether or not integer and abelian fractional QH transitions belong to the same universality class).  Thus, there is a need for alternate formulations of the QHIT, which can more easily address such questions.  	
	
	In this letter we present a first step in devising alternate formulations of the QHIT, making use of a dual composite fermion (CF) representation,  building on pioneering ideas of flux attachment,\cite{jain1989, Jainbook,zhang1989,Lopez91,Kalmeyer1992,Halperin1993,PasquierHaldane,Read1998} and particle-vortex duality\cite{Son2015,Seiberg:2016gmd}.
	As we show below, in a mean-field approximation, the CF formulation of the IQHIT belongs to the same universality class as the one studied in electron coordinates.  However, it offers several distinct advantages:  most interestingly, delocalized states occur over all energies\cite{Kumarsusy,Kumar2019b} at the IQHIT in the CF representation enabling a finite dc conductivity as $T \rightarrow 0$.    Furthermore, a CF theory can more readily incorporate interaction effects, and can treat integer and fractional QHITs on equal footing\cite{Kumar2020}.  
	
	The phase diagram of the IQHIT is realized in the CF representation as follows: first, the integer QH state of electrons with $\sigma_{xy} = e^2/h$ maps onto an integer QH state of CFs but with opposite Hall conductivity.  Second, the electron insulator is a CF insulator.   It only remains to show that the critical exponents obtained in the CF representation are identical to those predicted by the CCM.

	Using a tight-binding regularization of a CF hamiltonian, we compute two critical exponents, $\nu$ and $\eta$ describing respectively the divergence of the localization length, and wavefunction multifractalilty.   We find $\nu = 2.56 \pm 0.02$, and $\eta = 0.51\pm 0.01$, both of which are in excellent agreement with established results obtained from the CCM. \cite{Lee1993, Slevin2009, Obuse2010, Amado2011, Fulga2011, Slevin2012, Obuse2012, Nuding2015, Puschmann2019, Sbierski2020, Gruzberg2020, Evers2008a} 
	Thus, we establish that the IQHIT as viewed in CF coordinates is governed by the same fixed point  as the CCM.  This observation opens new possibilities in studies of the IQHIT, where interaction effects may be included more readily.  	
	
	\noindent\textit{ IQHIT in the idealized CF model -} 2D electrons in a perpendicular magnetic field $B$ can be transformed, via an exact mapping (``flux attachment")\cite{zhang1989, Lopez91} to CFs that couple to the sum of the external and ``statistical" flux $B + b(r)$.\cite{Kalmeyer1992,Halperin1993}  When two quanta of flux are attached to each electron, there is an exact identity relating the CF density to the statistical flux: $b(r) = -4 \pi n(r)$.  CF mean-field theory results from ``smearing the flux" and the identity is satisfied only on average: $\langle b(r) \rangle = -4 \pi \langle n(r) \rangle$.  With a quenched random potential $V(r)$ that varies on length scales large compared to the magnetic length, the linear response is a random density $\delta n(r) = \chi V(r)$, where $\chi = m/2\pi$ is the uniform compressibility \footnote{The mass $m$ that enters the compressibility is the same as the band mass, which is renormalized by Coulomb interactions.  However, no further renormalization occurs due to gauge fluctuations, as shown in Ref. \onlinecite{KimFurusakiWenLee1994}}.  Thus, in CF mean-field theory, there is a slaving\cite{Wang2017,Kumarsusy} between $V(r)$ and $b(r)$:\footnote{From here on, we change our notation so that $b(r)$ and $a(r)$ represent the ``effective'' magnetic and gauge fields experienced by the composite fermions. The effective field is related by a shift to the ``statistical'' gauge field, i.e. $a_{\rm eff.}(r) = a_{\rm stat.}(r) - A(r)$.}  $V(r) = - b(r)/2m$.  
	Furthermore, for asymptotic behavior near criticality, we can ignore non-linear response effects, and study the following model Hamiltonian density:
	\begin{equation}
	\label{idealham}
	h(r) =  c^{\dagger} \left[ \frac{\left(-i \bm \partial - \bm a \right)^2}{2m} + \frac{g}{2} \frac{b(r)}{2m} - \mu \right] c,  \ \ b(r) = \epsilon_{ij} \partial_i a_j.
	\end{equation}
	It involves free,  spin-polarized fermions with parabolic dispersion coupled to a random vector potential $\bm a(r)$, along with a ``gyromagnetic term" $\frac{g}{2} b(r)/2m$.    In the above context of CF mean-field theory with long-wavelength disorder, $g=2$.

Surprisingly, when $g=2$,  the system undergoes an IQHIT as the spatial-average magnetic field $b_0 \equiv \overline b(r)$ changes sign.  To see why, observe that when $b_0 \ne 0$, all finite energy states are localized in the thermodynamic limit at $T=0$.  However, there are  exact {\it zero energy modes}\cite{Aharonov1979} for $b_0 < 0$, which behave as a filled Landau level.  The zero modes are absent for $b_0 > 0$.  As a consequence,  the zero temperature phases are IQH (insulating) states  for $b_0 < 0$($b_0 > 0$).  At the critical point ($b_0 = 0$), the Hall conductivity can be computed {\it analytically} for the Hamiltonian above, and is $\sigma_{xy} = -e^2/2h$.\footnote{We assume that all odd moments of $V(r)$ vanish.}  It follows from the contrapositive of Laughlin's gauge argument that states at all energies are {\it delocalized} when $b_0 = 0$.  
Our present goal is to obtain critical exponents associated with this transition using a lattice realization of the above problem.  
	
	\noindent\textit{ Lattice model - }
		\begin{figure}[h!]
		\centering
		\includegraphics[width=2in]{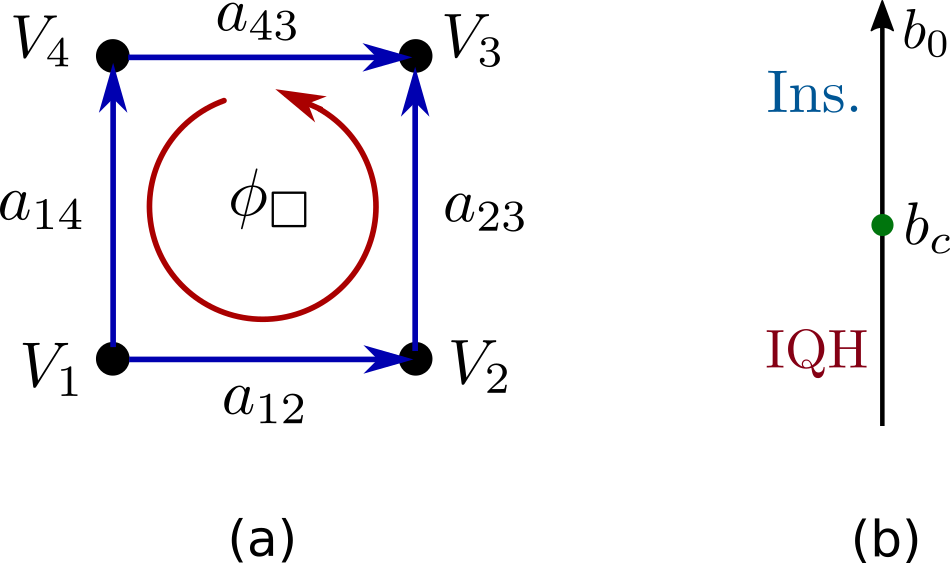}
		\caption{(a) Tight-binding model on a square plaquette (Eq. \ref{CF_lattice}). The flux  is proportional to the average potential on the attached vertices (see Eq. \ref{slaving_lattice}). (b) Phase diagram of the lattice model. In the idealized limit, a topological phase transition between a $\nu_{CF} = -1$ IQH state ($b_0 < 0$), and Insulator ($b_0>0$) occurs at $b_0 =b_c=0$.}
		\label{fig:lattice_model}
	\end{figure}
The  lattice  analog of the above consists of CFs on a square lattice with nearest neighbor hopping (we set the lattice spacing to unity).   Quenched random scalar and  vector potentials live on the lattice sites and  links 
respectively (Fig. \ref{fig:lattice_model}):
	\begin{equation}
	\mathcal{H}_{\rm lattice} = - \sum_{\langle i j \rangle} c^{\dagger}_{i} \left[ t e^{i a_{ij}}  + \mu \delta_{ij} \right]c_j  -\sum_{i} V_{i}c^\dagger_{i}c_{i},
	\label{CF_lattice}
	\end{equation}
	where $i, j$ label lattice sites, and $a_{ij}=-a_{ji}$ are associated with the directed nearest-neighbor link connecting sites $i$ and $j$.   Random fluxes are associated with each square plaquette of the lattice.  

	We slave the random chemical potential to the random flux as follows.  Consider a square plaquette, whose vertices are lattice sites labeled $1-4$ in a counterclockwise sense (Fig. \ref{fig:lattice_model}). We equate the flux asscociated with the plaquette, 
	\begin{equation}
	\phi_{\square} =  a_{12} + a_{23} + a_{34} + a_{41},
	\end{equation}
	with the {\it average} of the 4 random potentials $V_i$:
	\begin{equation}
	\phi_{\square} = b_{\square} = -\frac{m}{g}\sum_{i=1}^4 V_i,\label{slaving_lattice}
	\end{equation}
where, for simplicity, we take the mass to be the effective mass of the clean tight-binding model at the bottom of the band, i.e. $m = 1/2t$. 
We repeat the procedure for all elementary plaquettes of the lattice. We choose $V_i, i \in 1 \cdots 4$ from an independent uniform distribution  $V_{i} \in [-W/2+V_0, W/2+V_0]$, where $V_0 = -gb_0/4m$ and $W$ measures the strength of the disorder.

	\begin{figure*}
		\centering
			\includegraphics[width=5.5in]{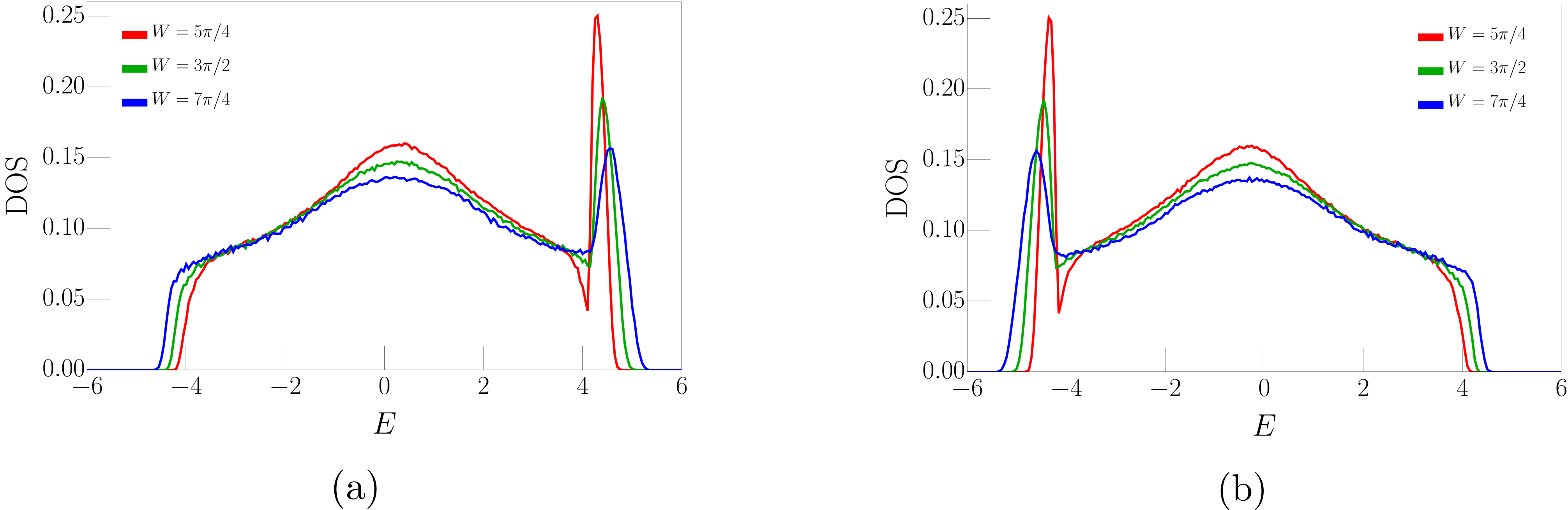}
		\caption{Density of states of the Hamiltonian \eqref{CF_lattice} for (a) $b_0 = 0.5$ (b) $b_0 = -0.5$. There are zero-modes near the bottom of the band for $b_0 < 0$ that become sharper as disorder strength is reduced. Further, they levitate  for $b_0 > 0$.}
		\label{fig:lattice_DoS_zero_modes}
	\end{figure*}

Therefore, for a weak, long range disorder and Fermi energy close to the bottom of the band, the Hamiltonian in Eq. \ref{CF_lattice} can be approximated as:
	\begin{gather}
	\mathcal{H}_{\rm lattice} \approx \frac{(\bm p - \bm a)^2}{2m} + \frac{g}{2}\frac{b(\bm r)}{2m} - 4t.
	\end{gather}
	%
	Notice that since the flux through each plaquette is bounded in magnitude by $\pi$, we have $|W/2 \pm V_0| \leq g\pi/4m$.  In principle one could adopt a more sophisticated procedure whereby the compressibiilty is determined in a self-consistent manner in equating the potential and flux disorders.  We choose not to do so for simplicity: as we show below the simple procedure employed here is already sufficient to capture the universal properties associated with the critical point, provided the Fermi energy remains sufficiently close to the band-bottom to warrant an effective mass approximation.

	Figure \ref{fig:lattice_DoS_zero_modes} displays the density of states over the entire bandwidth of the lattice model above, for non-zero $b_0$.  The lattice model increasingly accurately captures the behavior of the ideal model above in the limit where the Fermi level is close to the band bottom and the disorder is weak. For practical numerical calculations, however, it will be useful to use strong disorder which allows for shorter localization lengths and hence for better finite-size scaling behavior near the critical point.  This deviation from weak and long-wavelength disorder, as well as lattice corrections to the effective mass approximation, lead to a shift in the location of the phase transition as a function of $b_0$: in general, the IQHIT occurs at a finite value of $b_0$, which approaches zero as the idealized limit of the previous section is approached.

\noindent \textit{Localization length exponent - } Employing the standard transfer matrix techniques,\cite{MacKinnon1981, MacKinnon1983} we study the behavior of the localization length in the CF model above. We realize the tight-binding model on a quasi-1D cylinder of dimensions $L\times M$, where $L$ is the length of the cylinder along its axis while $M$ is the circumference. We obtain the localization length $\xi_M(b_0)$ along the axis of the cylinder  as a function of $b_0$ and the system width $M$ with $g=2$. In the 2D limit, i.e. $M \rightarrow \infty$, it diverges as $\xi_{\infty} (b_0) \sim |b_0-b_c|^{-\nu}$ near the critical point with the critical exponent $\nu$. We obtain $\nu$ via the finite-size scaling of the dimensionless localization length: $\Lambda_M(b_0) \equiv \xi_M(b_0)/M$ near the critical point.\cite{Slevin1999} To achieve this, we fit our data to the following polynomial function:
	\begin{align}
	\Lambda_M(b_0) &= \sum_{n=0}^{N_R} a_n \left( M^{1/\nu}\Delta\right)^n + \psi M^{-y} + c_{11} \psi \Delta M^{1/\nu} M^{-y}\label{fitting_function}
	\end{align}
	where $N_R$ is the degree of the polynomial in the relevant parameter $\Delta \equiv b_0-b_c$. $\psi$ is the amplitude of the leading irrelevant operator and $y$ is the corresponding correction to scaling exponent. Further, $a_n, c_{11}$ and $b_c$ are fitting parameters, the last of which gives the location of the transition.
	
	\begin{figure*}
		\centering
		\includegraphics[width=5.5in]{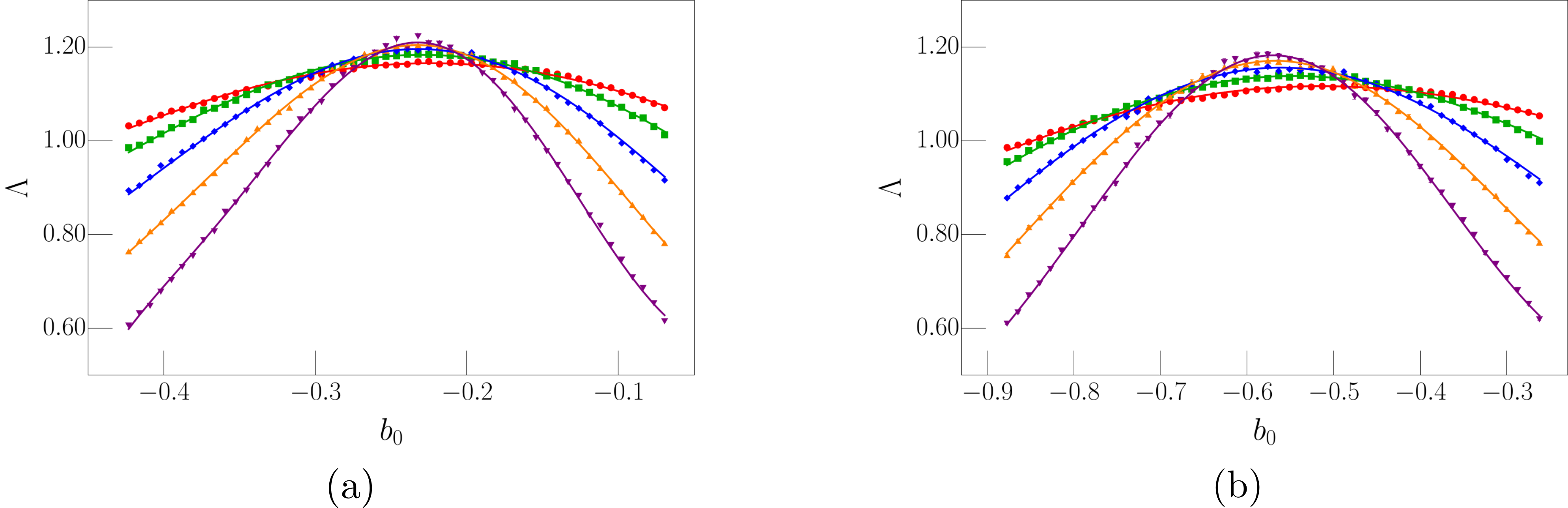}
		\caption{Scaling of the renormalized localization length as a function of $b_0$ at $E_F = -4$ and (a) $W=3\pi/2$, (b) $W=7\pi/4$. We use $L=10^7$ and the red to purple datapoints correspond to $M=16,32,64,128,256$ in order. The best fit to Eq. \eqref{fitting_function} is drawn with solid lines and we obtain the critical exponent: (a) $\nu=2.56 \pm 0.02$, (b) $\nu=2.57\pm 0.02$. Critical points are located at $b_0=b_c$ where (a) $b_c=-0.229$ and (b) $b_c = -0.558$.}
		\label{fig:loc_length_b0}
	\end{figure*}		
	For $W=3\pi/2$ and Fermi energy $E_F=-4$ at $g=2$, we plot the calculated $\Lambda_M(b_0)$ in Fig. \ref{fig:loc_length_b0}(a).  Fitting the data to the above polynomial form using the standard least square error method, we extract $\nu =2.56\pm 0.02$.  Also, for a stronger disorder: $W=7\pi/4$, we find $\nu=2.57 \pm 0.02$ (Fig. \ref{fig:loc_length_b0}(b)) suggesting that the exponent is independent of disorder strength. These results are in agreement with the previous studies of the IQHIT using the CCM.\cite{chalkercoddington, Lee1993, Slevin2009, Obuse2010, Amado2011, Fulga2011, Slevin2012, Obuse2012, Nuding2015, Puschmann2019, Sbierski2020, Gruzberg2020} They support the idea that the two descriptions of IQHIT lead to the same universal behavior. 
    In addition, we note that our results are slightly inconsistent with studies based on other models reporting a  smaller exponent. \cite{Gruzberg2017,Ippoliti2018,Zhu2019}

	\noindent \textit{Multifractal scaling - } In addition to the localization length exponent $\nu$, wavefunction multifractality  represent additional universal characteristics of the IQHIT. They correspond to the finite size scaling of the inverse participation ratios $P_q$ calculated from the critical wavefunction $\psi$:
	\begin{align}
	    P_q \equiv L^d\langle |\psi|^{2q} \rangle \propto L^{-2(q-1)-\Delta(q)}.\label{mf_fss}
	\end{align} 
	where $L$ is the system size and $d=2$. Employing standard techniques,\cite{RevModPhys.67.357, Evers2008b} we calculate these exponents using the critical wavefunctions of a square system of dimensions $L\times L$ with periodic boundary conditions. Since the total flux through the sample is quantized in the units of $2\pi$, we round $b_c$ obtained in the previous section to the nearest integer multiple of $2\pi/L^2$.

	\begin{figure*}[ht!]
		\centering
		\includegraphics[width=0.85\textwidth]{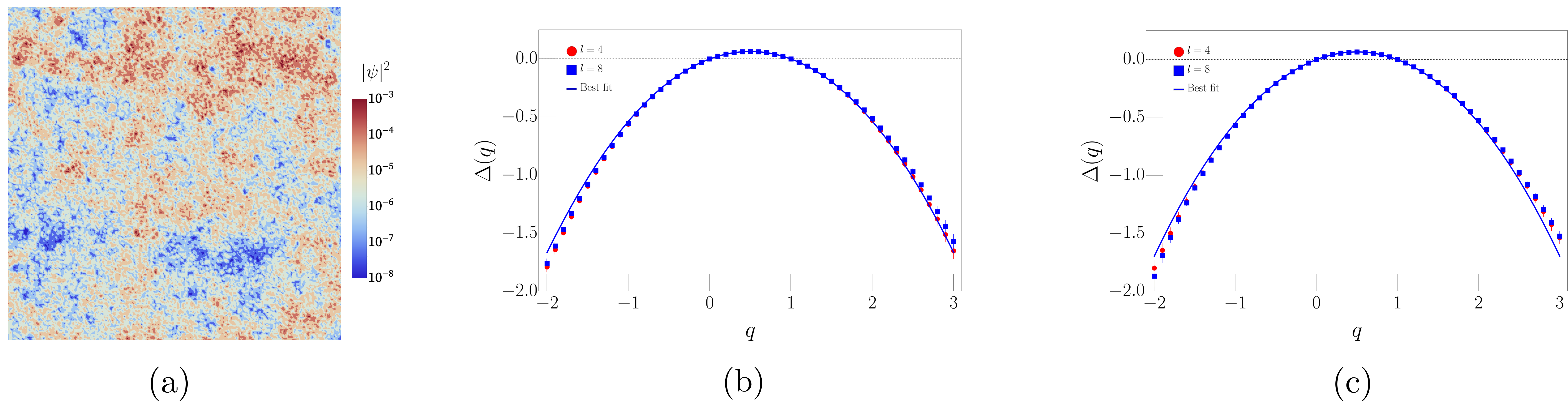}
		\caption{(a) A critical wavefunction $\psi$ displaying multifractal behavior. Numerically calculated $\Delta(q)$ at the IQHIT for $E_F = -4$ and (b) $W=3\pi/2$ and (c) $W = 7\pi/4$. We first average $|\psi|^2$ over a box of dimensions $l\times l$ and then over 1000 wavefunctions. Using $L=32, 64, 128$ and $256$,  $\Delta(q)$ is obtained by performing finite size scaling according to Eq. \eqref{mf_fss}. The best fit to Eq. \eqref{multifrac_fitting} using $l=8$ data is drawn with the solid blue line and we get (a) $\gamma=0.129 \pm 0.005$, (b) $\gamma = 0.133 \pm 0.006$.}
		\label{fig:multifractality}
	\end{figure*}
	
	For the critical point in Fig. \ref{fig:loc_length_b0}(a) at $b_0 = -0.229$, we find $\eta \equiv -\Delta(2) = 0.51\pm 0.01$. And for the critical point in Fig. \ref{fig:loc_length_b0}(b) at $b_0=-0.558$, we get $\eta = 0.52 \pm0.01$. These are close to the value $\eta = 0.5425$ obtained in Ref. \onlinecite{Evers2008a}. Further, they are also consistent with $\eta = 0.5$ predicted in Ref. \onlinecite{Zirnbauer2019}. We plot the full multifractal spectra in Fig. \ref{fig:multifractality} and fit them to the following form symmetric around $q=1/2$:\cite{Mirlin2006,Evers2008a}
	\begin{align}
	\Delta(q)=2q(1-q)\left[\gamma_0+\gamma_1(q-1/2)^2+\gamma_2(q-1/2)^2\right].
	\label{multifrac_fitting}
	\end{align}
	We find $\gamma_0 = 0.129 \pm 0.005, \gamma_1=0.003 \pm 0.003, \gamma_2=-0.0002 \pm 0.0004$ and $\gamma_0 = 0.133 \pm 0.006, \gamma_1=0.002 \pm 0.004, \gamma_2=-0.00005 \pm 0.00050$ for the two critical points. These are in excellent agreement with the corresponding quantities in Ref. \onlinecite{Evers2008a}.
	Likewise, we also find evidence for corrections to the proposed parabolic form\cite{Zirnbauer1999, Bhaseen2000, Tsvelik2007, Bondesan2017} since $\gamma_1 \neq 0$. It should be noted that our data does not show a perfect symmetry around $q=1/2$. We believe that this is due to finite size effects. Similar to Ref. \onlinecite{Evers2008a}, as we report in Appendix \ref{rq}, the asymmetry in $\Delta(q)$ approaches zero in the thermodynamic limit.
	
		\begin{table*}[ht!]
		\centering
		\begin{tabular}{
				>{\centering\arraybackslash} m{3cm}|
				>{\centering\arraybackslash} m{2cm}|
				>{\centering\arraybackslash} m{2cm}|
				>{\centering\arraybackslash} m{2.5cm}
		}
			\hline
			Parameters & $\nu$ & $\eta$ & $\gamma_0$ \\\hline
			$W=3\pi/2$ $b_0 = -0.229$ & $2.56\pm 0.02$ & $0.51\pm 0.01$ & $ 0.129 \pm 0.005$\\\hline
			$W=7\pi/4$ $b_0=-0.558$ &  $2.57\pm 0.02$ & $0.52\pm 0.01$ & $ 0.133 \pm 0.006$\\\hline
		\end{tabular}
		\caption{Summary of exponents at Fermi energy $E_F = -4$ and $g=2$.}
		\label{tab:results}
	\end{table*}
	We summarize the results of all obtained critical exponents in Table \ref{tab:results}. 
	
	While the value $g=2$ in Eq. \ref{idealham} is motivated by CF mean-field theory, we can consider the effect of relaxing the value of $g$ on the IQHIT.  Such deviations from $g=2$ can arise from lattice corrections to the effective mass approximation, or from the breaking of particle-hole symmetry in the disorder-averaged theory\footnote{For Particle-hole (PH) symmetry at $\nu=1/2$, the Hall conductivity of CFs at $b_0=0$ must be $\sigma_{xy}^{\rm CF} = -e^2/2h$. This was shown to be the case for $g=2$ in Eq. \ref{idealham} in Ref.  \onlinecite{Kumarsusy}. However, if $g\neq 2$, the PH-symmetry can be broken.}.  As we show in Appendix \ref{generalized_model}, the localization length exponent decreases monotonically  with $g$. The extent to which such deviations\cite{Klumper2019}  reflect a new universality class for the IQHITs, or are due to substantial finite size effects, or from large corrections to scaling from irrelevant operators,  remain unclear and require further study.  We shall return to these questions in future work.

	\noindent \textit{Discussion - }Our results have several important implications for the IQHIT, and suggest several new directions of exploration.  The most important implication of our study governs finite temperature dc transport in the quantum critical regime.   In electron coordinates, extended states occur at a single energy, and without any interaction effects, $\rho_{xx}(T \rightarrow 0) \ne \rho_{xx}(T = 0)$.  By contrast, in the CF representation, this issue does not arise, since extended states occur over a range of energies at criticality.  Indeed, a finite CF resistivity implies the same for the electrical resistivity via the exact relation\cite{Kivelson1992}
	\begin{equation}
	\rho_{\rm CF}^{ab} = \rho_{\rm el}^{ab} + 4 \pi \epsilon^{ab}, \ \ \epsilon^{ab} = \left(\begin{array}{cc} 0 & 1 \\ -1 & 0 \end{array} \right).
	\end{equation}
	It is thus the CF representation that guarantees a smooth $T\rightarrow 0$ limit of the resistivity tensor in mean-field theory.  
	
	Second, the success of the CF mean-field theory suggests new analytic approaches to describing the non-interacting IQHIT.  Recent work has shown that the effective theory governing disorder averaged quantities in the weak-coupling regime $\sigma^{\rm CF}_{xx} \gg 1$, is a non-linear sigma model with a topological term, similar to the theory put forward in electron coordinates\cite{Kumar2020}.  However, such theories run to strong coupling, since the critical point itself occurs at $\sigma_{xx}^{\rm CF} \sim \mathcal O(1)$.  
	Recently, a current algebra description of the IQHIT was proposed in Ref. \onlinecite{Zirnbauer2019}.  Our multifractal scaling results are in excellent agreement with the predictions of Ref. \onlinecite{Zirnbauer2019}.  However, the prediction for the localization length exponent in Ref. \onlinecite{Zirnbauer2019} requires much larger system sizes than our current simulations.
	It is likely that the CF representation may give way to new analytic treatments.  One possible route is to note that the theory in Eq. \ref{idealham} is equivalent to a 2-component Dirac fermion at finite chemical potential in the presence of a random vector potential.  The non-abelian bosonization of the Dirac fermion may lead to complementary descriptions in terms of Wess-Zumino-Witten models.  We shall report progress on such analytic treatments in  future studies.

	\noindent \textit{Conclusions - } In summary, we have calculated the critical and multifractal exponents for the IQHIT using a  composite-fermion representation, which are in agreement with numerical studies of the CCM.  While the electron and CF formulations have distinct origins, they are expected to flow to the same IR fixed point governing the IQHIT: in this sense, the electron and CF formulations are thus `dual' to one another.

	\begin{acknowledgments} 
		We thank S. Kivelson, J.-H. Son, and M. Zirnbauer for fruitful discussions. K. S. H. was supported in part by the Department of Physics, Stanford University through an undergraduate summer research fellowship. S. R. and P. K. were supported in part by the U.S. Department of Energy, Office of Basic Energy Sciences, Division of Materials Sciences and Engineering, under Contract No. DE-AC02-76SF00515. P. K. was supported in part by DOE Grant No. DE-SC0002140.
	\end{acknowledgments}

\bibliography{bigbib}

\appendix

\section{Asymmetry in the multifractal spectrum $\Delta(q)$ \label{rq}}
\begin{figure}[ht!]
	\centering
	\includegraphics[width=0.3\textwidth]{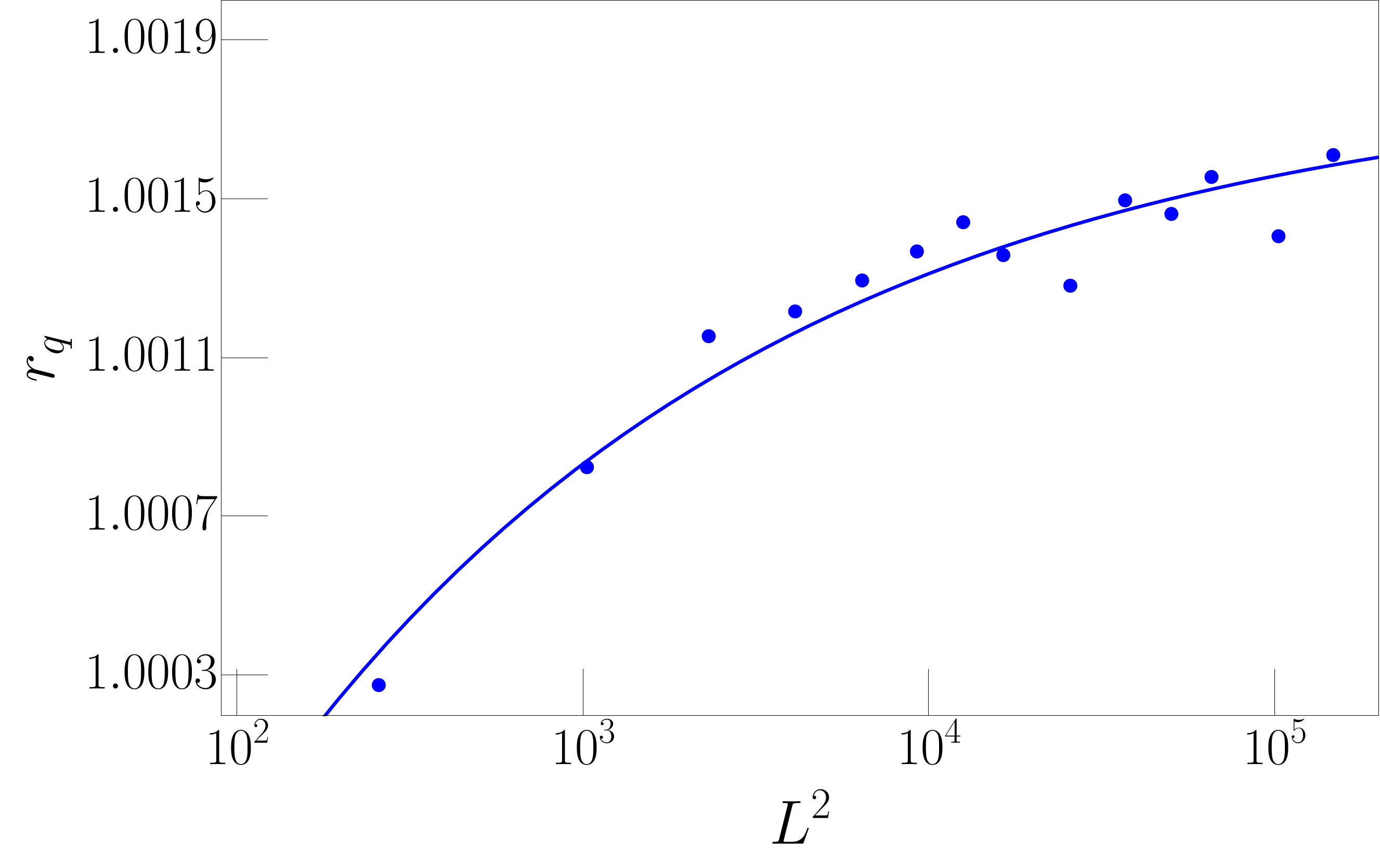}
	
	(a)
	\vspace{10pt}

	\includegraphics[width=0.3\textwidth]{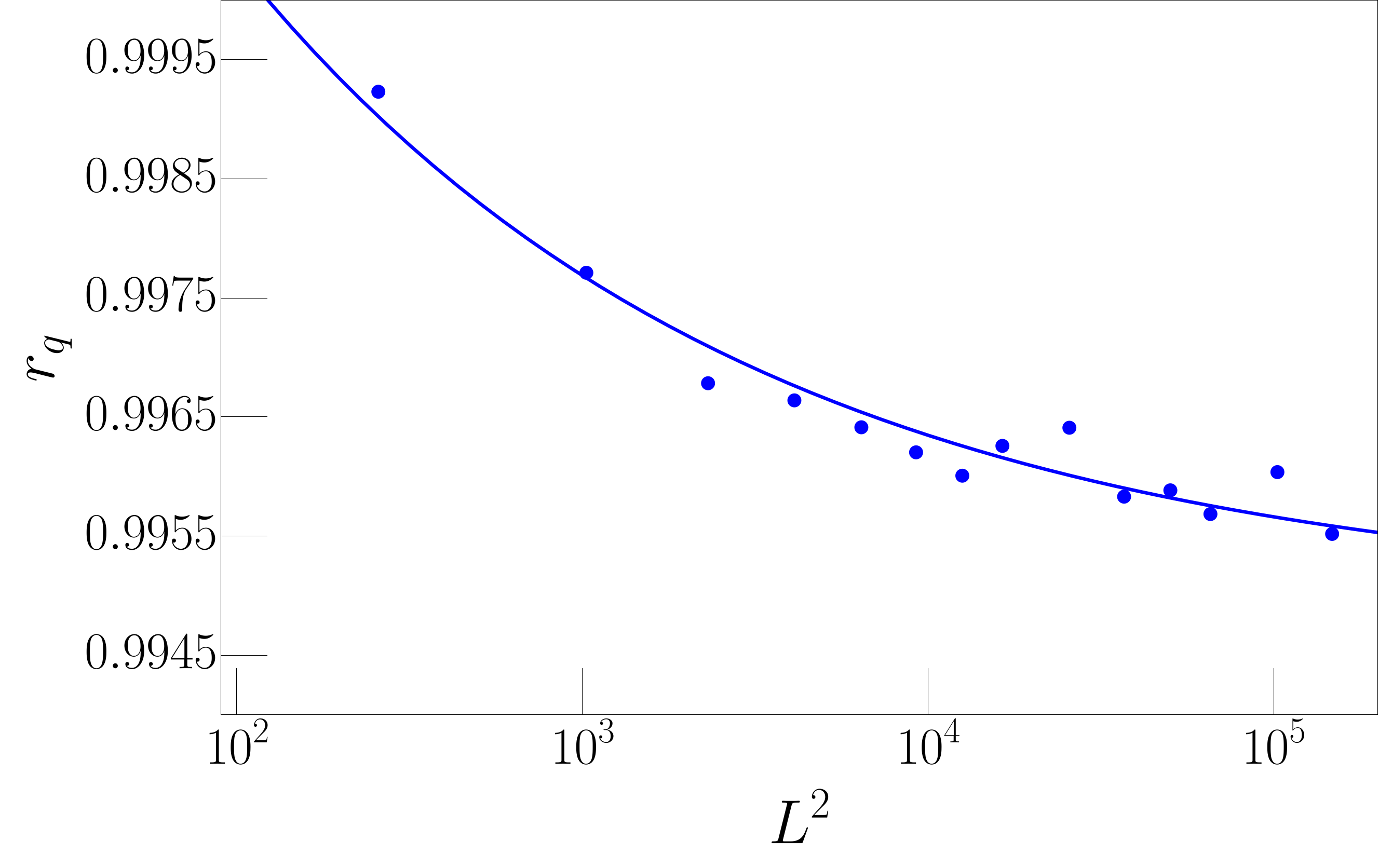}
	
	(b)
	\vspace{10pt}
	
	\includegraphics[width=0.3\textwidth]{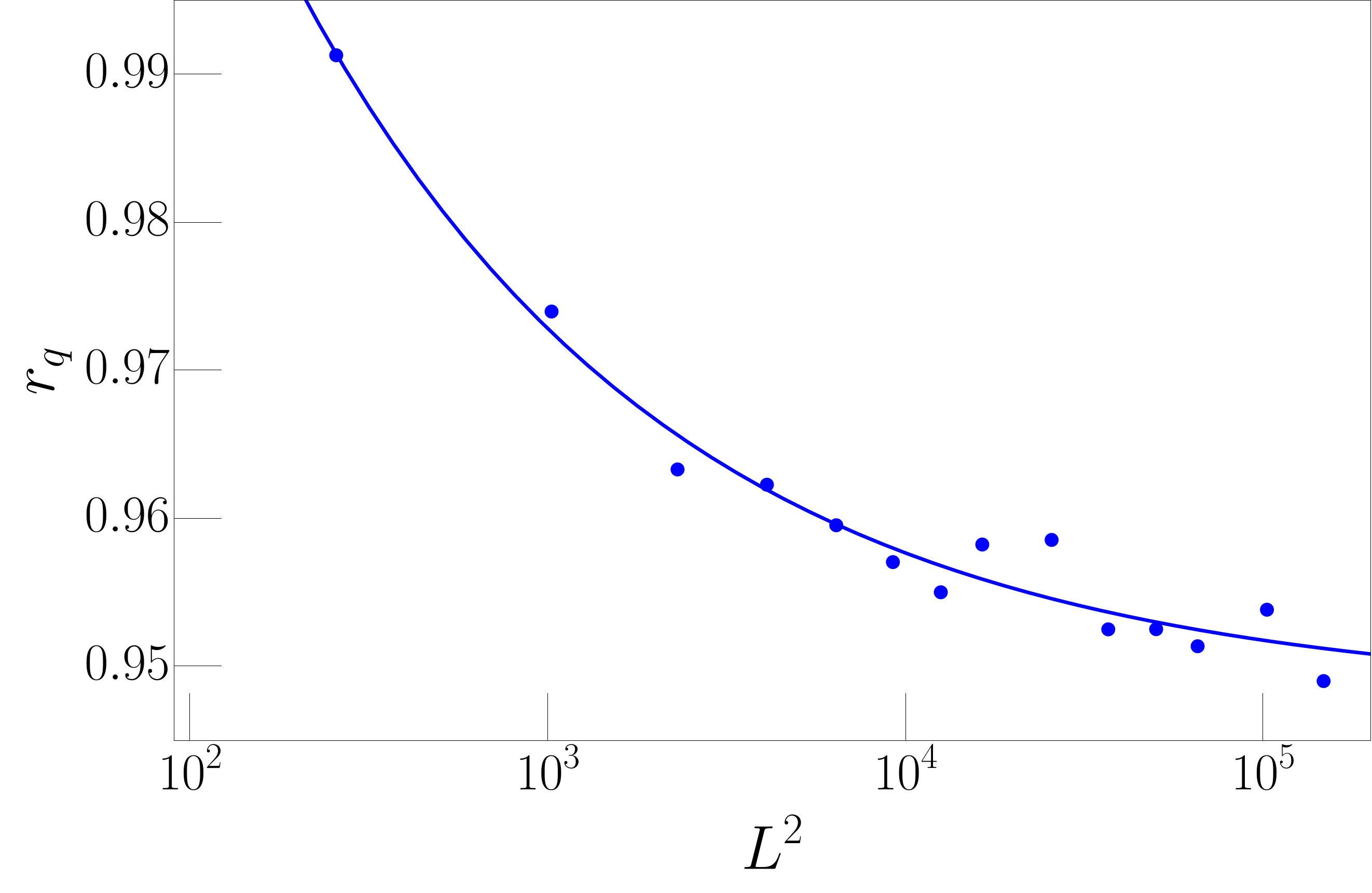}
	
	(c)
	\vspace{10pt}
	
	\caption{The finite size scaling behavior of $r_q$ at Fermi energy $E_F = -4$, $b_0=-0.229$ and disorder strength $W=3\pi/2$ for (a) $q=0.6$ (b) $q=1.1$ (c) $q=1.5$. 
		We use box size of $l=8$ and average over 1000 wavefunctions for each size. The solid curves are obtained by fitting the data to $r_q = a_1 L^{-x} + r_{\infty,q}$ using the least square error method. We find $r_{\infty, q} =$ (a) $1.0018$, (b) $0.9950$ and, (c) $0.9481$. These numbers are close to typical values obtained at finite sizes and thus support that the finite size effects are responsible for the asymmetry observed in the numerically calculated $\Delta(q)$.
	}
	\label{fig:rq}
\end{figure}

In this section, we report on the asymmetry of the multifractal spectrum $\Delta(q)$ with increasing system size. Following Ref. \onlinecite{Evers2008a}, we define:
\begin{gather}
	r_q \equiv L^{2(2q-1)}\frac{P_q}{P_{1-q}} \sim L^{-\Delta(q)+\Delta(1-q)}\tag{S1}\label{rq_eq}
\end{gather}
Therefore, $r_q$ saturates as a function of system size if $\Delta(q) = \Delta(1-q)$. We plot $r_q$ in Fig. \ref{fig:rq} for the critical point at $W = 3\pi/2$, $E_F = -4$ and $b_0 = -0.229$. Similar to Ref. \onlinecite{Evers2008a}, $r_q$ shows a tendency towards saturation as $L$ is increased.

\section{Critical scaling in a generalized model \label{generalized_model}}
In the main text, we looked at the IQHIT induced by tuning $b_0$ at $g=2$ in the CF lattice model of Eq. 2. This was motivated by the fact that $b_0$ is a physical tuning parameter corresponding to the deviation of electron filling fraction away from $\nu=1/2$. As mentioned in the main text, deviation away from $g=2$ can characterize effects such as broken particle-hole symmetry or corrections to the effective mass approximation. 
Motivated by this, we have studied the phase diagram of the lattice model in the two parameter space of $b_0$ and $g$ using the transfer-matrix approach.

\begin{figure}
	\vspace{25pt}
	\centering
	\includegraphics[width=3in]{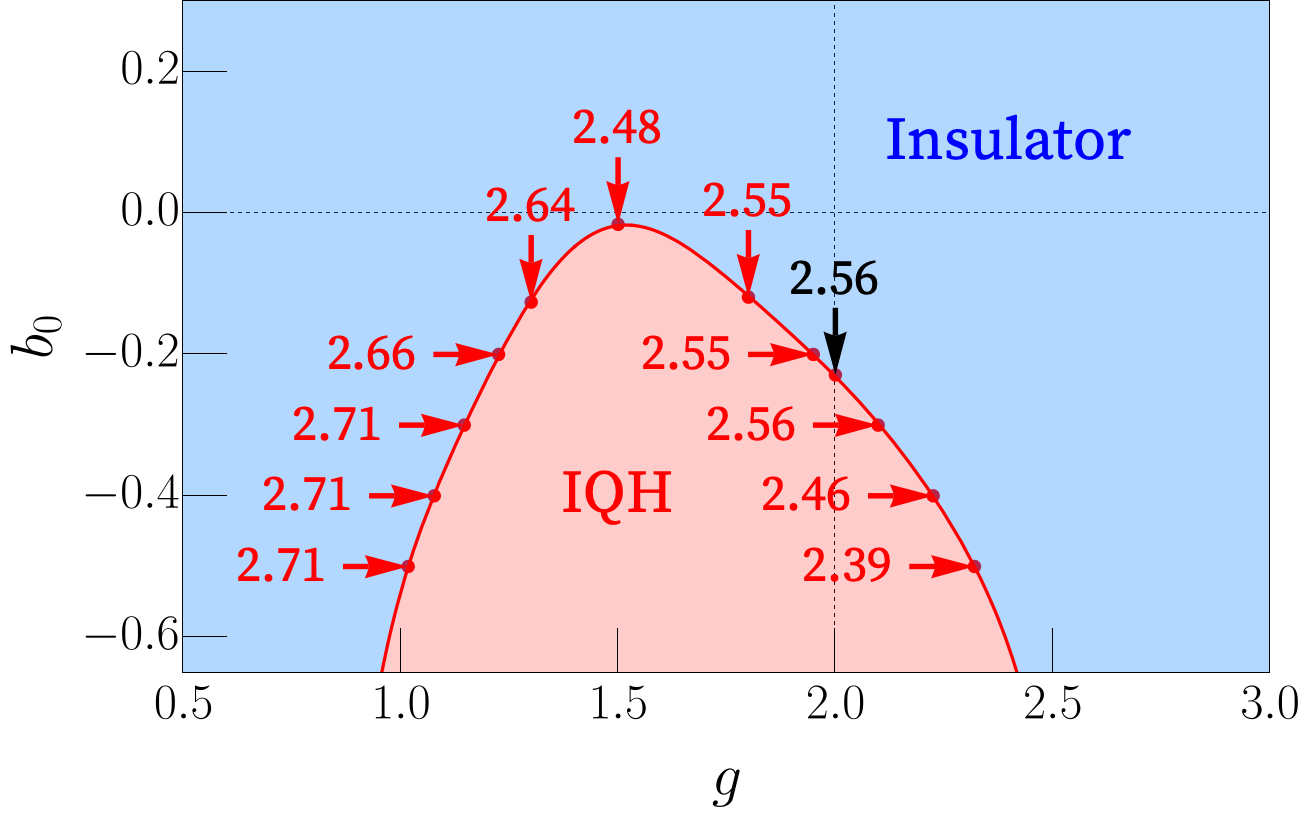}
	\caption{The phase diagram and critical exponents $\nu$ of the lattice model in the two parameter space of the average magnetic field $b_0$ and the gyromagnetic ratio $g$. The parameters are $E_F = -4$, $W = 3\pi g/4$. The IQH state with $\nu_{\rm CF}=-1$ is separated from the insulator through the critical curve drawn in red. The arrows show the critical exponents calculated at various points on this curve. The results suggest that $\nu$ decreases as $g$ increases and thus the IQHITs are not in the same universality class. However, a detailed study involving bigger system sizes is needed to show that finite size-effects are not responsible for such a behavior.}
	\label{fig:exponents_curve}
\end{figure}

\begin{figure*}
	\centering
	\begin{minipage}{0.33\textwidth}
		\includegraphics[width=0.9\textwidth]{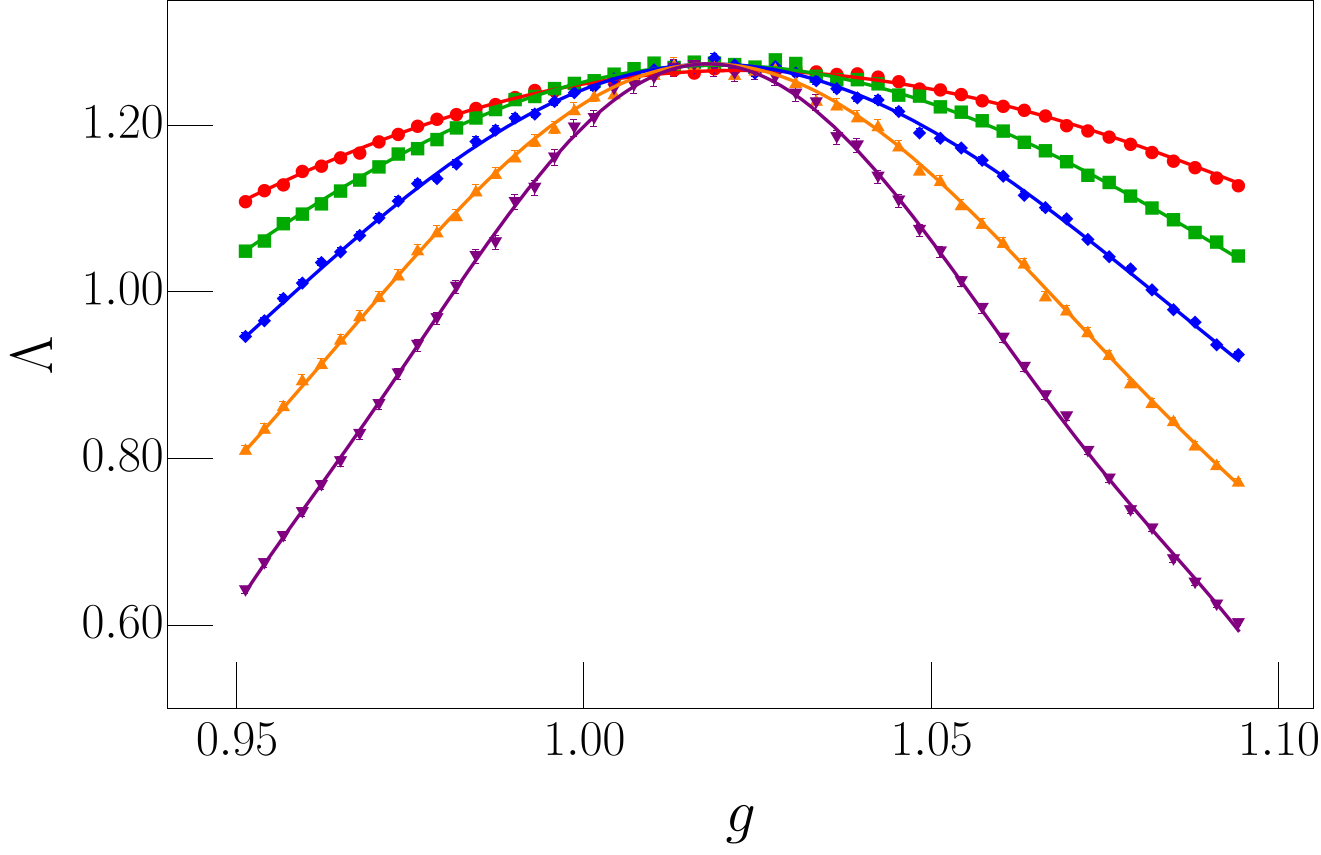}\\\vspace{5pt}
		(a) $b_0=-0.5,\ g=1.017,\ \nu=2.71$
	\end{minipage}%
	\begin{minipage}{0.33\textwidth}
		\includegraphics[width=0.9\textwidth]{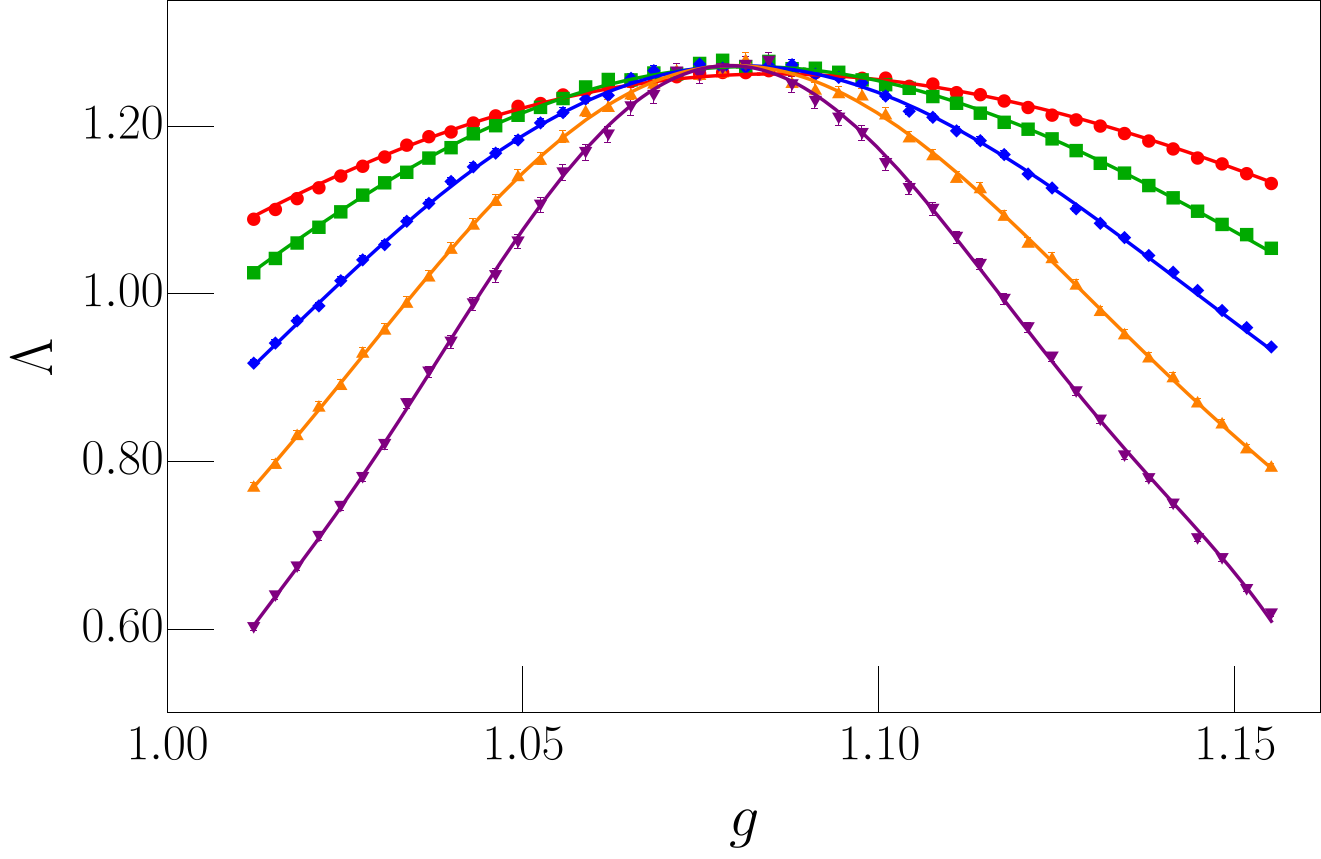}\\\vspace{5pt}
		(b) $b_0=-0.4,\ g=1.077,\ \nu=2.71$
	\end{minipage}%
	\begin{minipage}{0.33\textwidth}
		\includegraphics[width=0.9\textwidth]{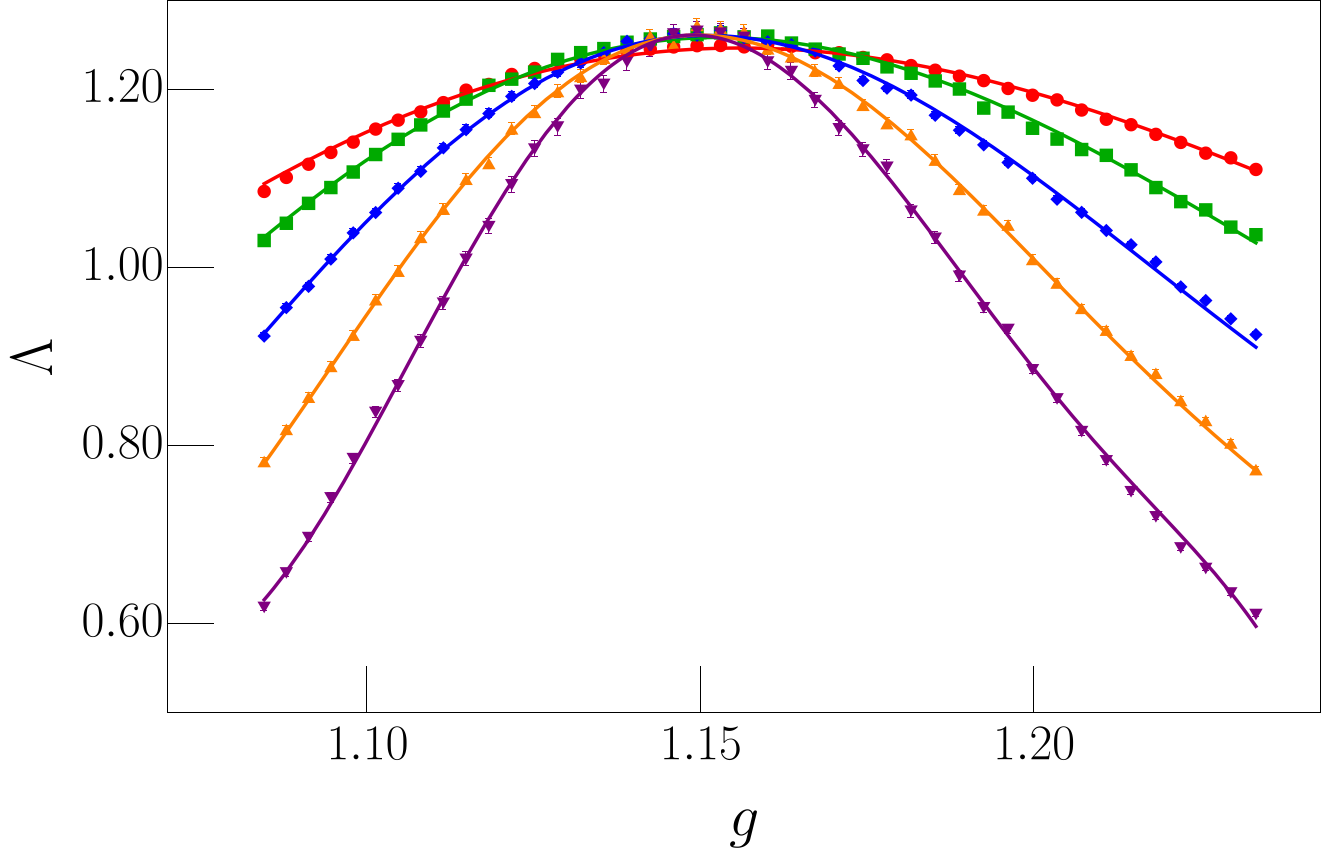}\\\vspace{5pt}
		(c) $b_0=-0.3,\ g=1.146,\ \nu=2.71$
	\end{minipage}\vspace{15pt}
	
	\begin{minipage}{0.33\textwidth}
		\includegraphics[width=0.9\textwidth]{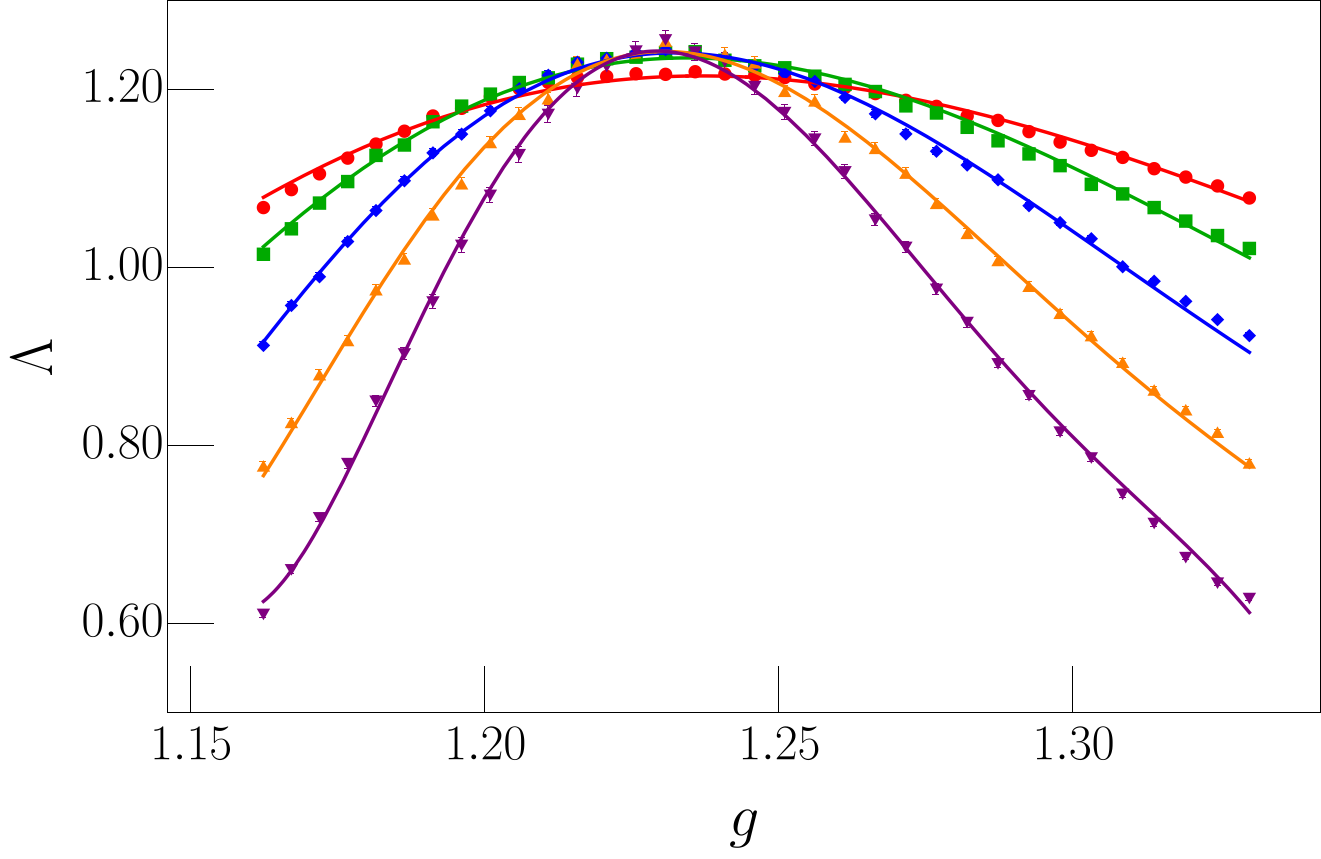}\\\vspace{5pt}
		(d) $b_0=-0.2,\ g=1.225,\ \nu=2.66$
	\end{minipage}%
	\begin{minipage}{0.33\textwidth}
		\includegraphics[width=0.9\textwidth]{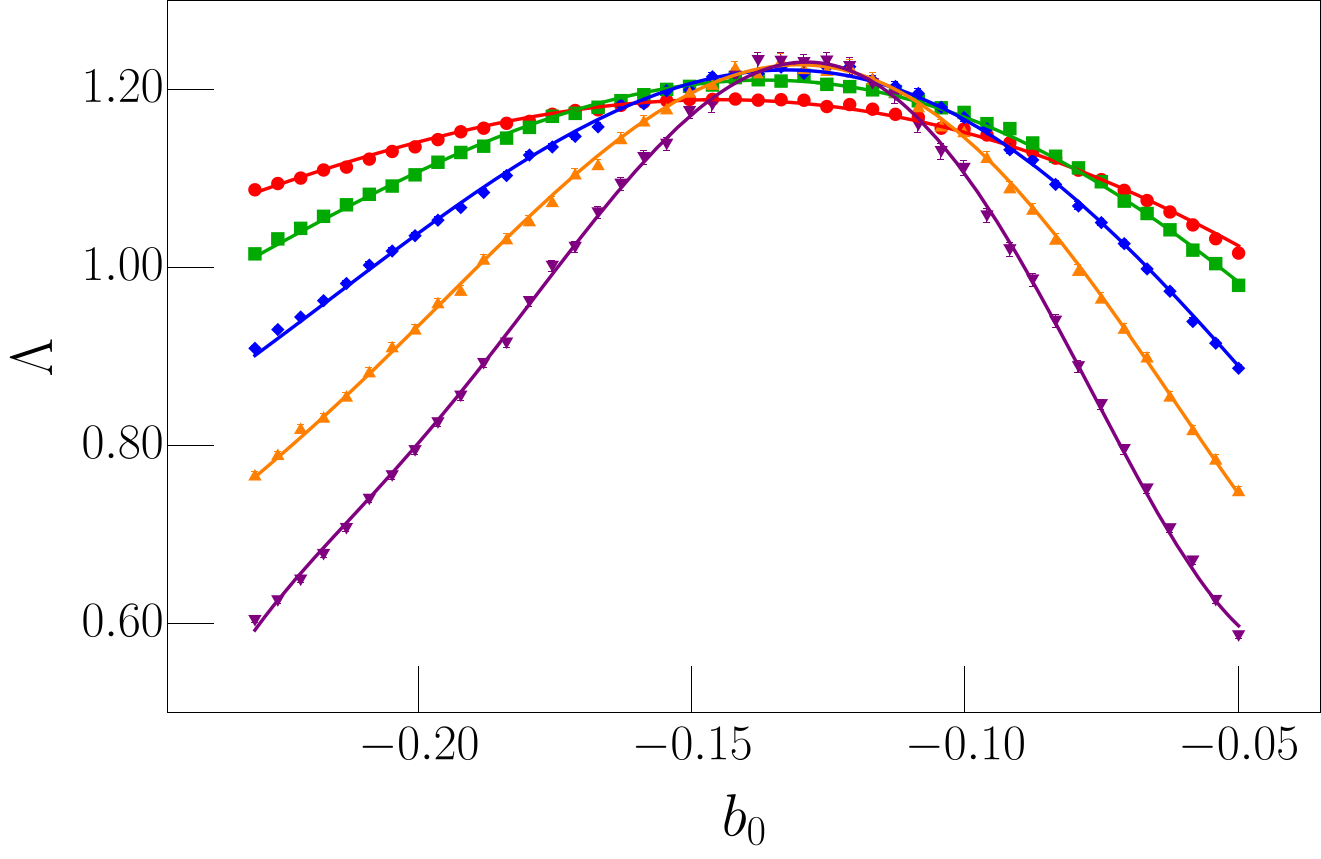}\\\vspace{5pt}
		(e) $b_0=-0.126,\ g=1.3,\ \nu=2.64$
	\end{minipage}%
	\begin{minipage}{0.33\textwidth}
		\includegraphics[width=0.9\textwidth]{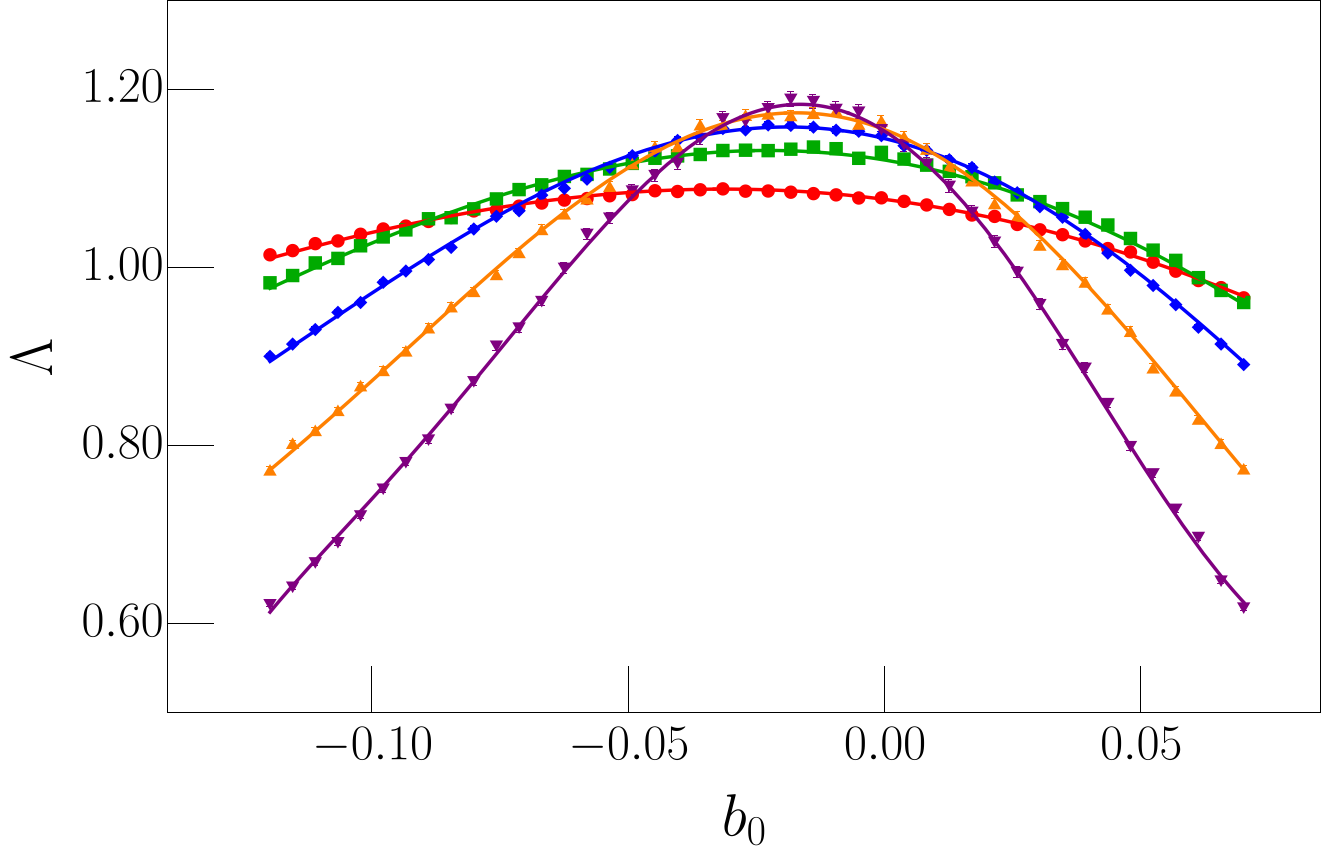}\\\vspace{5pt}
		(f) $b_0=-0.016,\ g=1.5,\ \nu=2.48$
	\end{minipage}\vspace{15pt}
	
	\begin{minipage}{0.33\textwidth}
		\includegraphics[width=0.9\textwidth]{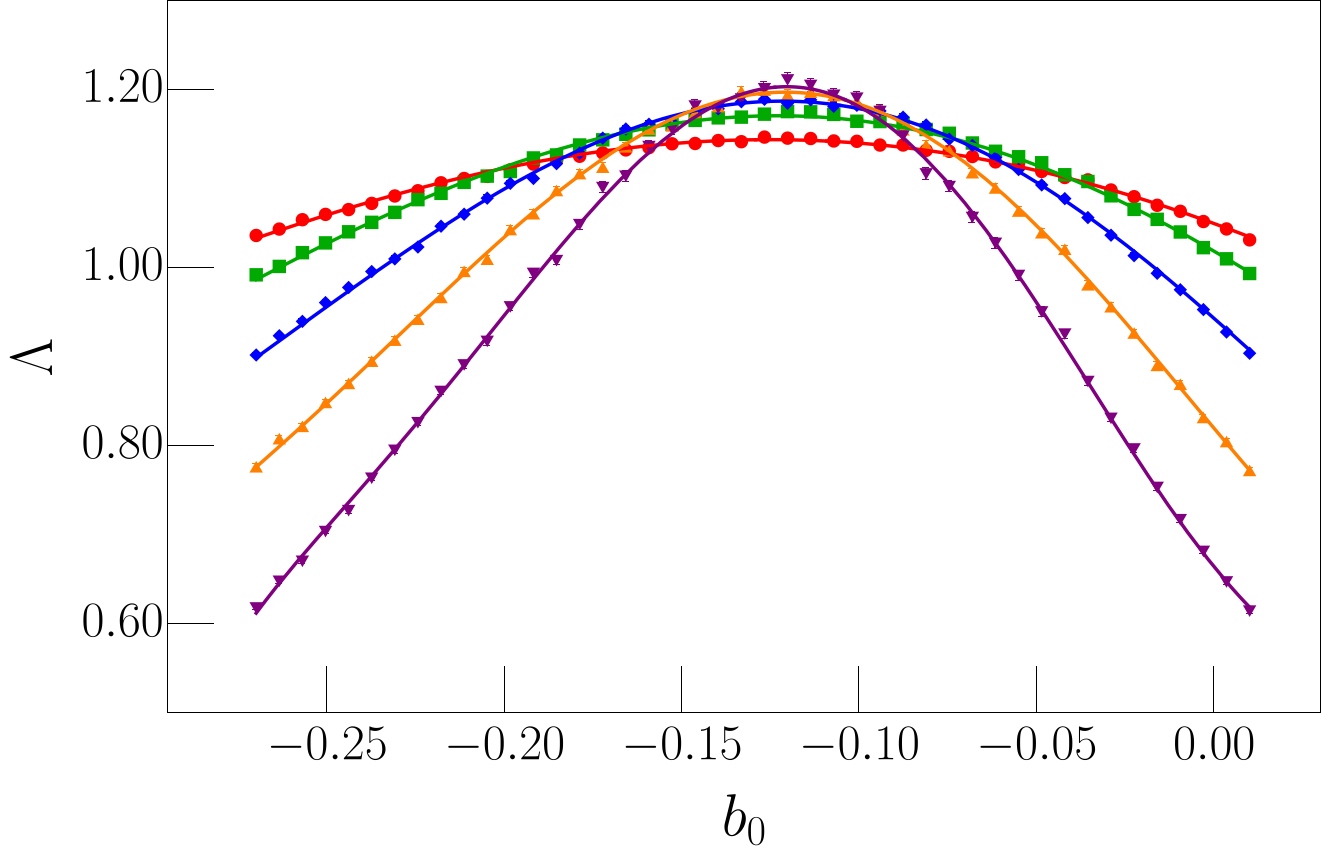}\\\vspace{5pt}
		(g) $b_0=-0.119,\ g=1.8,\ \nu=2.55$
	\end{minipage}%
	\begin{minipage}{0.33\textwidth}
		\includegraphics[width=0.9\textwidth]{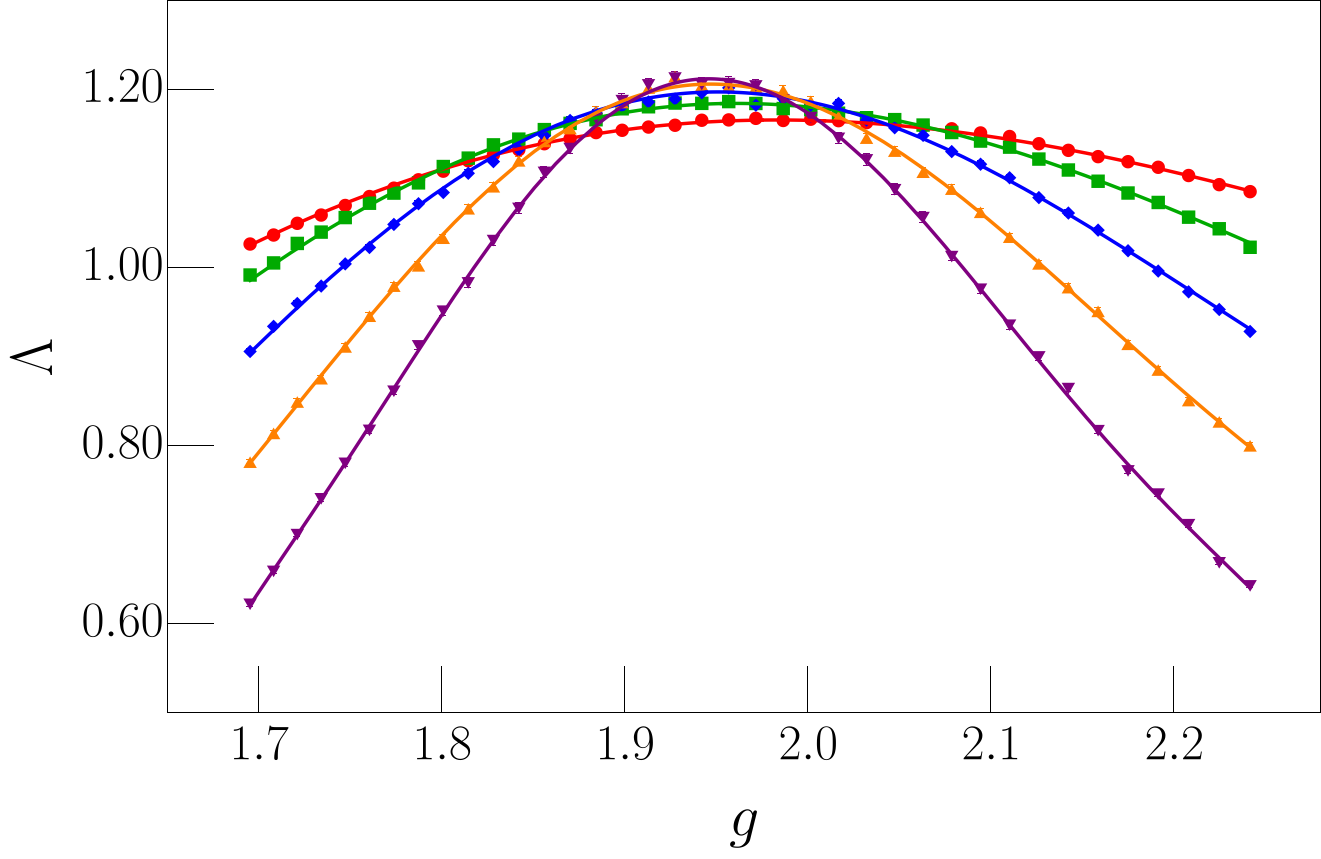}\\\vspace{5pt}
		(h) $b_0=-0.2,\ g=1.948,\ \nu=2.55$
	\end{minipage}%
	\begin{minipage}{0.33\textwidth}
		\includegraphics[width=0.9\textwidth]{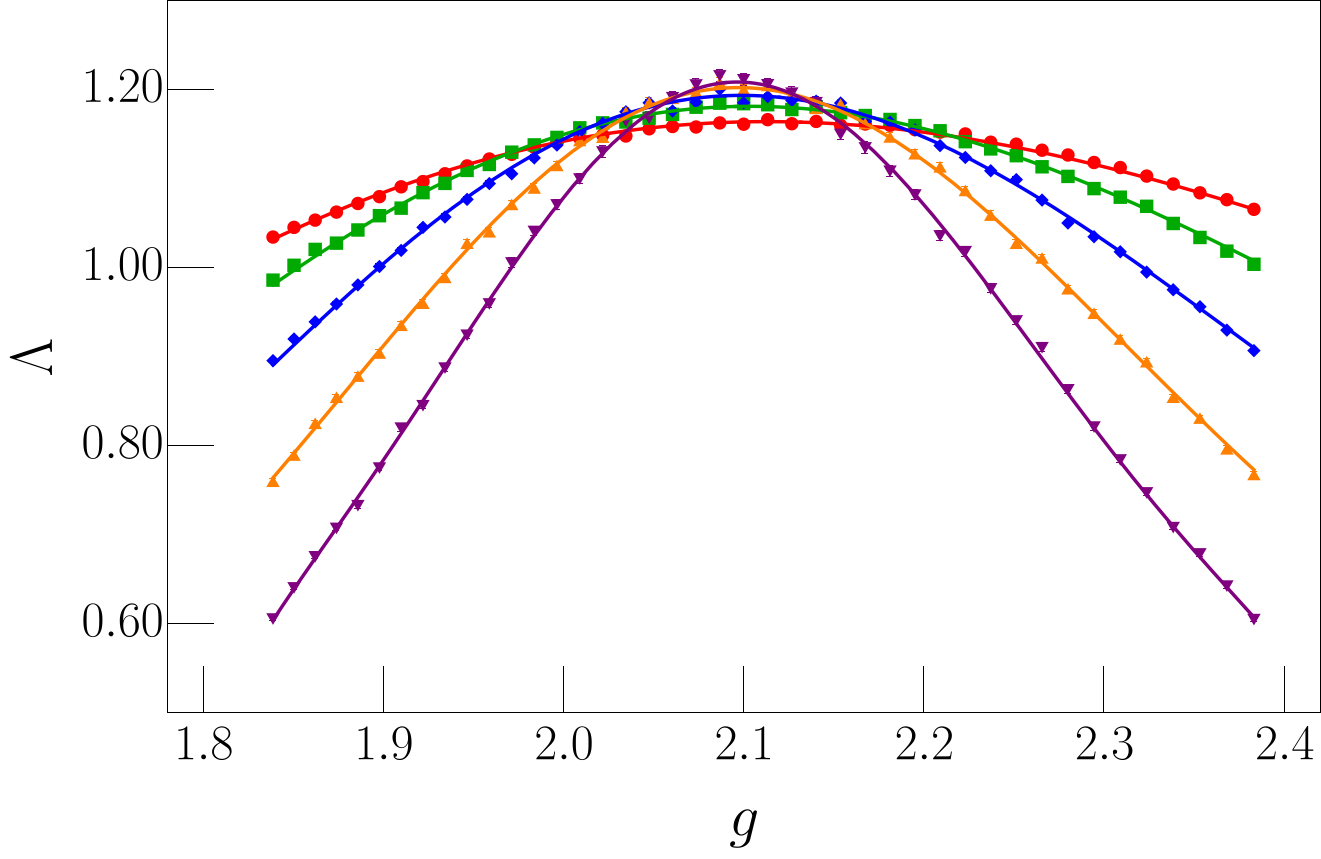}\\\vspace{5pt}
		(i) $b_0=-0.3,\ g=2.098,\ \nu=2.56$
	\end{minipage}\vspace{15pt}
	
	\begin{minipage}{0.33\textwidth}
		\includegraphics[width=0.9\textwidth]{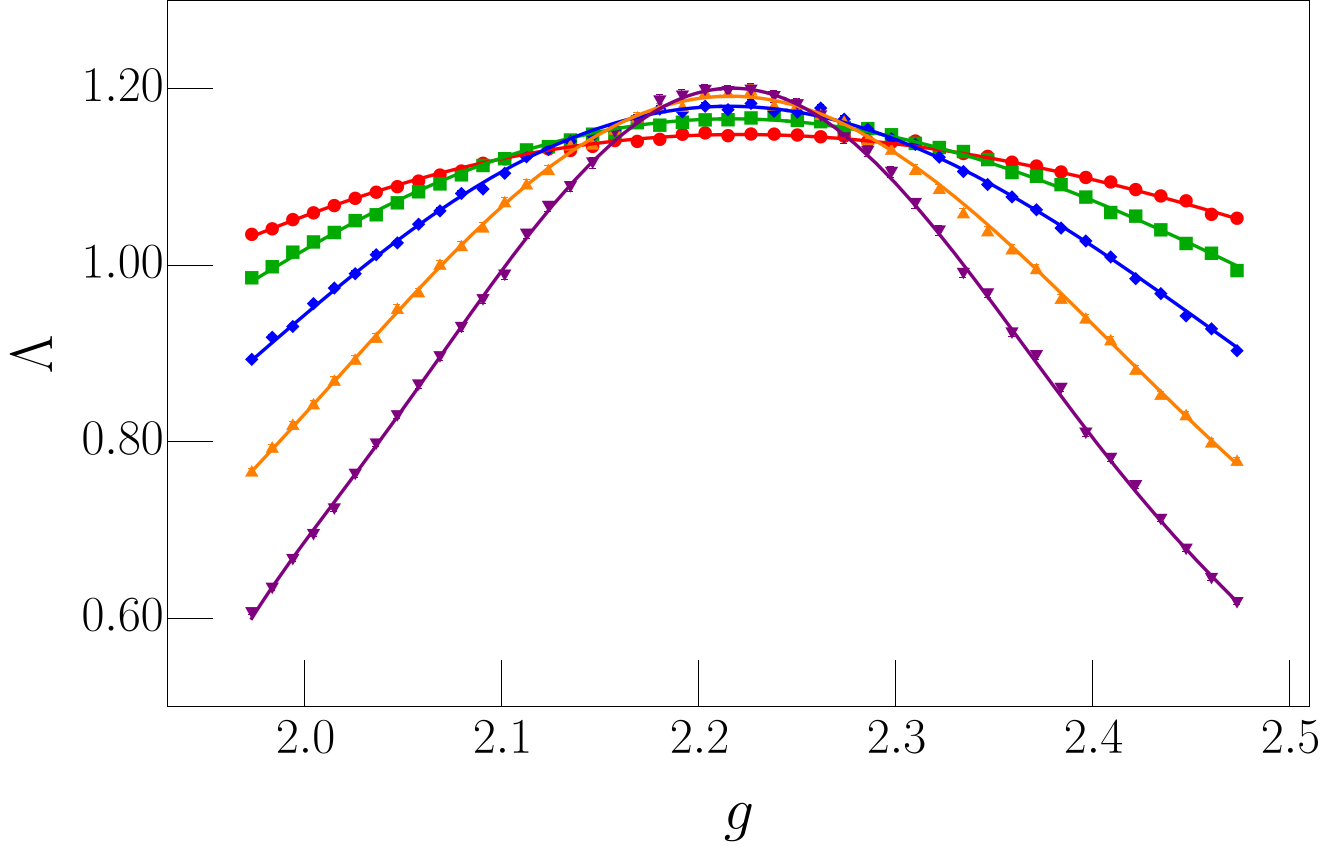}\\\vspace{5pt}
		(j) $b_0=-0.4,\ g=2.226,\ \nu=2.46$
	\end{minipage}%
	\begin{minipage}{0.33\textwidth}
		\includegraphics[width=0.9\textwidth]{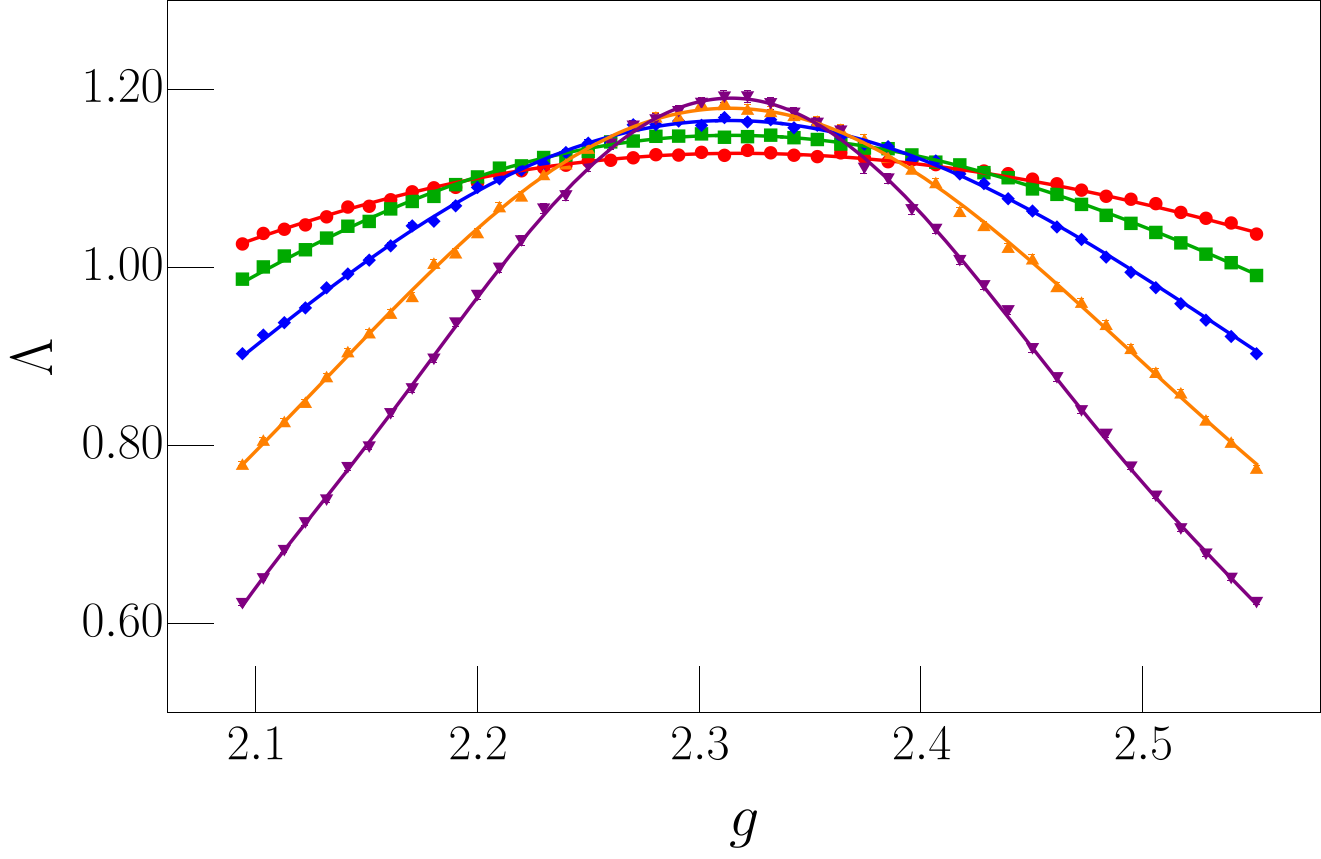}\\\vspace{5pt}
		(k) $b_0=-0.5,\ g=2.319,\ \nu=2.39$
	\end{minipage}
	\caption{Localization length vs. tuning parameter for the critical points shown in Fig. \ref{fig:exponents_curve}. The tuning parameter is taken to be $g$ for the subfigures (a)-(d) and (h)-(k), while it is $b_0$ for others. The red dots, green squares, blue rotated-squares, orange triangles, purple upside-down-triangles correspond to $M=16,32,64,128,256$ respectively. We have set $L=10^7$ for all the data points. Further, $W = 3\pi g/ 4$ and $E_F = -4$. The data around each critical point is fitted to Eq. 6 of the main text to obtain the location of the transition and the critical exponent $\nu$. The errors in the critical exponents are approximately $0.02$.}
	
	\label{fig:data_exponents_cruve}
\end{figure*}

In the two-parameter space of $g$ and $b_0$, the IQH phase is separated from the insulating phase by a critical curve. We first obtain the approximate critical curve by calculating $\Lambda_M$ as a function of $g$ for various slices of the two-parameter space with constant $b_0$. The critical curve intersects a constant $b_0$ line at zero-dimensional critical points. These critical points can be located using the fact that at these points $\Lambda_M$ becomes either a constant or a slightly increasing function of $M$. Then, by performing a detailed finite-size scaling analysis using Eq. 8 in the vicinity of these points, we obtain both the critical exponent and the precise shape of the critical curve. The results are plotted in Fig. \ref{fig:exponents_curve}. In addition, we provide the data and best fits in Fig.  \ref{fig:data_exponents_cruve}.

As can be seen in Fig. \ref{fig:exponents_curve}, the critical exponent $\nu$ monotonically decreases as $g$ is increased. Therefore, all the IQHITs may not be in the same universality class and thus the critical exponent may change if Particle-hole symmetry in the electron problem is broken.  We note that Ref. \onlinecite{Klumper2019} also reported the presence of a fixed line in the geometrically disordered network model. These results are surprising because IQHITs are thought be described by Pruisken's non-linear sigma model\cite{Pruisken1984} and the two-parameter scaling\cite{Khmelnitskii1983, Pruisken1985} proposed in this context suggests that there is only one critical point. We think that due to the varying degrees of disorder strengths and finite-size corrections, more rigorous studies involving bigger system sizes and also different techniques are required to resolve the discrepancy between theory and numerics. Further analytic studies of g-2 corrections in this framework may help discriminate between finite size effects and an exact fixed line.

\newpage

\end{document}